\documentclass[reprint,aps,prl,amsmath,amssymb]{revtex4-1}%,superscriptaddress,linenumbers

\usepackage{lmodern}
\usepackage{graphicx}

\usepackage{hyperref}
\hypersetup{colorlinks=true,allcolors=blue,pdfstartview={FitH}}
\usepackage[all]{hypcap}

\newcommand{\tuwien}{Institute of Applied Physics, TU Wien, Wiedner Hauptstra{\ss}e 8-10/E134, 1040 Wien, Austria}

\DeclareRobustCommand{\woods}[3]{%
\mbox{$(#1\hspace{0.1em}\times\hspace{0.1em}#2)$}%
\if\relax\detokenize{#3}\relax%
\relax%
\else%
$R$#3$^\circ$%
\fi
}
\DeclareRobustCommand{\oneby}[1]{%
\woods{1}{1}{}%
\if\relax\detokenize{#1}\relax%
\relax%
\else%
\textsubscript{#1}
\fi
}
\DeclareRobustCommand{\twoby}{\woods{2}{1}{}}
\DeclareRobustCommand{\degC}{$^\circ$C}
\DeclareRobustCommand{\muO}{$\mu$\textsubscript{O}}
\newlength{\pagefigure}
\newlength{\columnfigure}
\DeclareRobustCommand{\floatref}[3][]{%
\hyperref[#2]{#3}~\ref{#2}%
\if\relax\detokenize{#1}\relax%
\relax%
\else%
\hyperref[#2]{(#1)}%
\fi}

\DeclareRobustCommand{\Fig}[2][]{\floatref[#1]{#2}{Fig.}}

\DeclareRobustCommand{\Figs}[1]{\hyperref[#1]{Figs.}~\ref{#1}}
\AtBeginDocument{%
    \newwrite\bibnotes
    \def\bibnotesext{Notes.bib}
    \immediate\openout\bibnotes=\jobname\bibnotesext
    \immediate\write\bibnotes{@CONTROL{REVTEX41Control}}
    \immediate\write\bibnotes{@CONTROL{%
    apsrev41Control,author="08",editor="1",pages="1",title="0",year="1"}}
     \if@filesw
     \immediate\write\@auxout{\string\citation{apsrev41Control}}%
    \fi
}%

\begin{document}

\title{2D Surface Phase Diagram of a Multicomponent Perovskite Oxide: \texorpdfstring{La$_{0.8}$Sr$_{0.2}$MnO$_3$}{La0.8Sr0.2MnO3}(110)}
\date{\today}

\author{Giada Franceschi}
\author{Michael Schmid}
\author{Ulrike Diebold}
\author{Michele Riva}
\email[Corresponding author: ]{riva@iap.tuwien.ac.at}
\affiliation{\tuwien}

\begin{abstract}
The many surface reconstructions of (110)-oriented lanthanum--strontium manganite (La$_{0.8}$Sr$_{0.2}$MnO$_3$, LSMO) were followed as a function of the oxygen chemical potential (\muO{}) and the surface cation composition. Decreasing \muO{} causes Mn to migrate across the surface, enforcing phase separation into A-site-rich areas and a variety of composition-related, structurally diverse B-site-rich reconstructions. The composition of these phase-separated structures was quantified with scanning tunneling microscopy (STM), and these results were used to build a 2D phase diagram of the LSMO(110) equilibrium surface structures.
\end{abstract}

\maketitle

The surfaces of oxide materials, both in their bulk form and as (ultra)thin films, are rarely bulk-terminated. Instead they form a multitude of diverse and complex structural configurations as a function of their oxygen content \cite{surnev2003vanadium,li2009two,sedona2005ultrathin,shaikhutdinov2012ultrathin,netzer2010small,shimizu2012effect}, their cation composition (for multielement materials) \cite{wang2016transition,franceschi2020atomically}, and, in case of metal-supported ultrathin films, the support itself \cite{sedona2005ultrathin,netzer2010small}. This manifold of phases is determined by a delicate interplay between polarity compensation \cite{noguera2000polar}, variability of the oxidation state of the metal atoms as a function of the oxygen chemical potential (1/2$\mu$\textsubscript{O$_2$} \cite{reuter2001composition,franceschi2019growth}, henceforth \muO{} for simplicity), strain and defect-formation energies \cite{wang2014vacancy}, and, depending on the preparation conditions, kinetic effects \cite{netzer2010small}.

\begin{figure*}[t]
\includegraphics[width=\pagefigure]{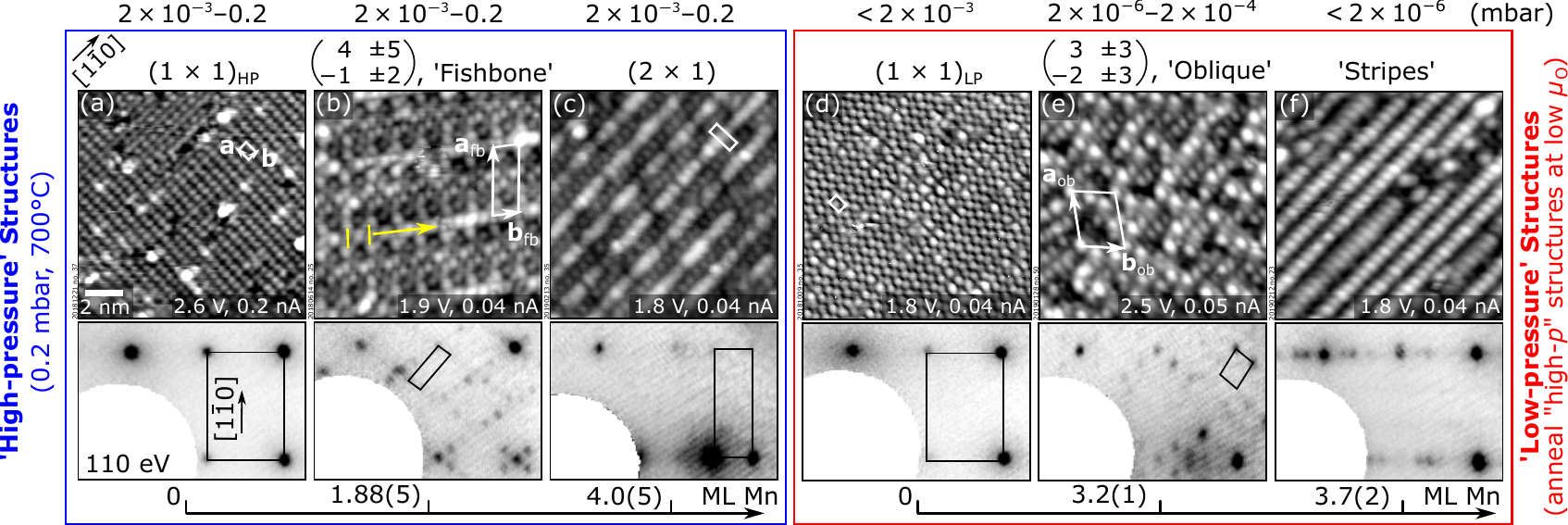}
\caption{\label{fig:1}Equilibrium surface structures of LSMO(110). Top: 12~$\times$~12~nm$^2$ STM images; bottom: LEED. O$_2$ pressure ranges (millibar) where each reconstruction is stable at 700\degC{} are reported at the top. (a--c) High-pressure (HP) structures, obtained by depositing Mn or La in PLD plus O$_2$ annealing \cite{franceschi2020atomically}. (d--f) Low-pressure (LP) structures, formed by annealing the HP structures at low pressure. The \oneby{HP/LP} surfaces have the same cation composition. Each structure is identified at the top by its superstructure periodicity and/or by a descriptive, short-hand label used in the text. Relative compositions are given in the bottom axes [1~ML is the number of Mn atoms in an (AMnO)$_2$ plane of LSMO(110), $4.64 \times 10^{14}$~cm$^{-2}$].}
\end{figure*}

% In catalysis and energy production applications, the material's activity may be ruled by the atomic structure \cite{riva2018influence}, composition, oxidation state of Mn species \cite{ponce2000surface,chen2015recent}, and surface oxygen vacancies \cite{}; in thin-film heterostructures, they determine electronic and magnetic properties \cite{huang2018atomically,de2005evidence}; it is also paramount for gaining fundamental insights about the LSMO electronic structure and excitations \cite{de2005evidence,krempasky2010bulk}. 

The surface structure of metal oxides dictates their surface properties through the coordination, arrangement, and electronic structure of the topmost atoms \cite{debenedetti2018atomic,mcbriarty2018potential,pan2020structural,jakub2019partially,jakub2019local,riva2018influence,zhu2016controlling,qin2020surface}. Hence, it is important to map out the parameter space of the equilibrium surface phases of these materials, including their evolution with \muO{}. Such a characterization has come a long way for binary oxides, but is still in its infancy for perovskite oxides (ABO$_3$), despite the undisputed role of their surfaces in many emerging and established technologies \cite{zubko2011,bhalla2000perovskite,pena2001chemical,kumah2020epitaxial}.

One prominent example is lanthanum--strontium manganite (LSMO). Owing to its many and diverse physical properties (among others, half metallicity, colossal magnetoresistance, metal--insulator transition, anti/ferromagnetic to paramagnetic transitions, high electronic conductivity at elevated temperatures), LSMO is used in a wide range of applications, e.g., spintronics \cite{majumdar2013pulsed,park1998direct,haghiri2004spintronics}, catalysis \cite{hwang2017perovskites,ponce2000surface}, energy production \cite{jiang2008development,ruiz2011symmetric}, and various thin-film technologies \cite{liao2019metal,huang2018multifunctional,huang2018atomically}. Because many relevant processes and interactions take place at the LSMO surfaces and interfaces, gaining a comprehensive understanding of the equilibrium surface phases of LSMO at the local (i.e., atomic) level is of paramount importance. Such investigations must be performed as a function of both, the cation, and the oxygen concentrations (the latter depends on the value of \muO{} at which the surface is treated): Both these parameters define a rich phase diagram including the physical properties named above \cite{hemberger2002structural}, the performance of LSMO-based devices \cite{kim2012probing,adler2004factors,bristowe2011surface,chen2015recent}, and emergent phenomena in LSMO thin films \cite{fan2019emergent}. %both of them determine the reactivity to specific reactions \cite{chen2015recent}, the performance of LSMO-based electrochemical \cite{kim2012probing,adler2004factors} and information-technology \cite{bristowe2011surface} devices, and emergent magnetic phenomena in LSMO thin films \cite{fan2019emergent}. %through changes, e.g., in density of oxygen vacancies \cite{kim2012probing} and the Mn oxidation state \cite{chen2015recent}.
To date, however, there exists no comprehensive investigation that simultaneously maps out the equilibrium surface phases of LSMO (nor of any other multielement oxide) as a function of the cation composition and \muO{}. The few studies available have unveiled that perovskite oxide surfaces typically consist of B-site-rich structures made of one or two atomic layers of differently linked polyhedra \cite{andersen2018pauling,enterkin2010homologous}, and have focused on their relation to the A:B cation ratio at the surface. In some cases, quantitative compositional relations between the different structures have been established \cite{wang2016transition,franceschi2020atomically,gerhold2014stoichiometry,feng2013reconstructions}. 

This work pushes the experimental characterization of perovskite oxide surfaces, by investigating the effect of the cation composition \textit{and} \muO{} on the local surface properties of (110)-oriented LSMO. Single-crystalline films of La$_{0.8}$Sr$_{0.2}$MnO$_3$ were grown on Nb-doped SrTiO$_3$(110) substrates by pulsed laser deposition (PLD) ($\approx$ 100~nm, 700\degC{}, 1 Hz, 2.2~J/cm$^2$, 4~$\times$~10$^{-2}$~mbar O$_2$) \cite{franceschi2020atomically}. Their surfaces were investigated in an ultra-high vacuum (UHV) surface science setup attached to the PLD chamber, equipped with scanning tunneling microscopy (STM), low-energy electron diffraction (LEED), x-ray photoelectron spectroscopy (XPS), and low-energy He$^+$-ions scattering (LEIS) (for details about the experimental methods, see Refs.~\onlinecite{franceschi2020atomically,gerhold2016adjusting}, and Section S1). Previous studies have shown that these films exhibit composition-related surface reconstructions at 700\degC{} and 0.2 mbar O$_2$, and have established the differences in cation coverages between them \cite{franceschi2020atomically}. These structures, here referred to as ``high-pressure'' (HP), are shown for reference at the left-hand side of \Fig{fig:1}, and are described in detail in Section S2. The current work is based on the behavior of the HP reconstructions over a wide range of \muO{} values (between $\approx-2.1$ eV and $\approx-1.4$ eV, corresponding to annealing between UHV and 0.2 mbar O$_2$ at 700\degC{}): When decreasing \muO{} (i.e., at more reducing conditions), new surface structures are formed, and a rich phase diagram emerges. These new structures, referred to as ``low-pressure'' (LP), are collected at the right-hand side of \Fig{fig:1}, and are described in detail in Section S3. Their relative compositions, as derived in this work, are reported in the bottom axis in terms of monolayers (ML) of Mn. 

After exploring in detail the behavior of the HP structures with decreasing \muO{} and the process of phase separation, this work will propose a novel, STM-based approach to organize the equilibrium surface phases of LSMO(110) (both HP and LP) in a quantitative diagram as a function of the cation composition and \muO{}.

The first HP structure whose behavior with \muO{} is considered is the A-site-rich \oneby{HP} of \Fig[a]{fig:1}: This structure stays unaltered upon annealing at 700\degC{} and 2~$\times$~10$^{-3}$ mbar $\leq$ $p$\textsubscript{O$_2$} $\leq$ 0.2~mbar, but transforms into the \oneby{LP} of \Fig[d]{fig:1} at lower pressures (down to UHV). The \oneby{LP} is characterized by the same periodicity as its HP counterpart, and the same cation composition [see LEIS data in Fig.~S1(b)]. However, it displays a glide plane that is not present in the \oneby{HP}, and possesses a smaller oxygen content [Fig.~S1(b)]. 

\begin{figure*}[t]
\includegraphics[width=\pagefigure]{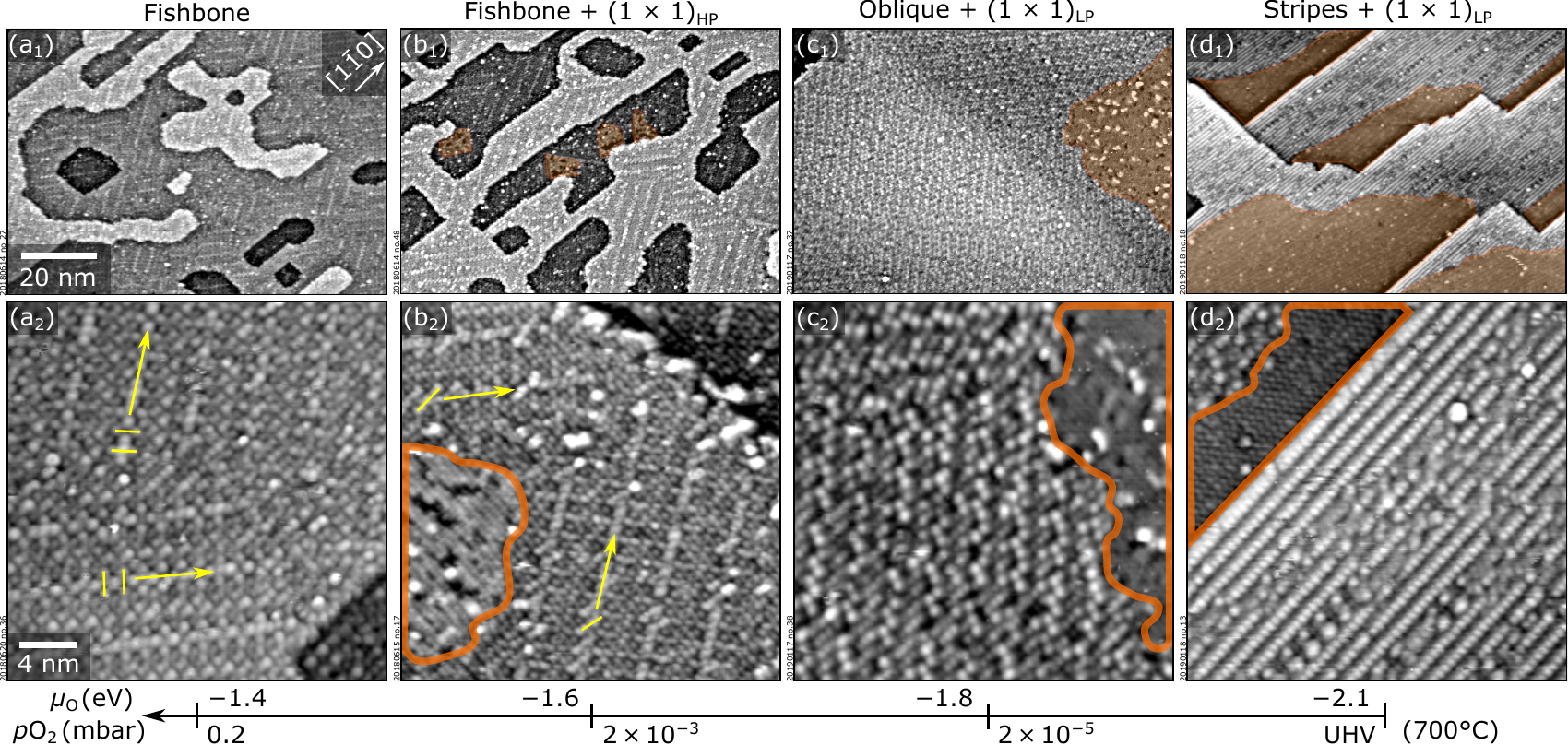}
\caption{\label{fig:2}Evolution of the fishbone surface upon annealing at increasingly reducing conditions. (a$_1$--d$_1$, a$_2$--d$_2$) STM images (100~$\times$~70 nm$^2$, and 26 $\times$ 26 nm$^2$, respectively). (a) Monophase fishbone surface. After annealing at $2 \times 10^{-3}$ mbar (b), small \oneby{HP} patches appear, and small structural changes occur in the fishbone. Annealing at $2 \times 10^{-5}$ mbar (c) produces \oneby{LP} areas, and transforms the remaining surface into the oblique structure of \Fig[e]{fig:1}. UHV annealing (d) enlarges the \oneby{LP} areas, and transforms the remaining surface into the stripes of \Fig[f]{fig:1}. The process is reversible.}
\end{figure*}

The other HP phases of LSMO(110) behave differently than the \oneby{HP} with decreasing \muO{}, as shown below. These structures are significantly Mn-richer and belong to a different family with respect to the \oneby{HP}, and are henceforth referred to as HP, `Mn-rich' phases. Notably, they continuously evolve into one other as a function of the Mn content, such that various surface structures exist between the fishbone and the \twoby{} of \Fig[b, c]{fig:1} \cite{franceschi2020atomically}. The qualitative behavior of these HP, Mn-rich phases with decreasing \muO{} is exemplified by \Fig{fig:2}, which shows the evolution of the fishbone phase of \Fig[b]{fig:1}, initially prepared at 0.2~mbar and 700\degC{}. When decreasing the O$_2$ pressure to 2~$\times$~10$^{-3}$~mbar at 700\degC{}, small \oneby{HP} areas appear (orange in \Fig{fig:2}). These patches become larger as \muO{} decreases, and change their atomic structure to \oneby{LP}. Meanwhile, the remaining fishbone-reconstructed surface undergoes a minor structural change: The small features highlighted by the short yellow lines in \Fig[a$_2$,~b$_2$]{fig:2} orient closer to the [1$\bar{1}$0] direction, an indication of a slight Mn enrichment \cite{franceschi2020atomically}. Between $1 \times 10^{-4}$ and $5 \times 10^{-6}$~mbar [\Fig[c]{fig:2}], larger \oneby{LP} patches form, while the remaining surface exposes the oblique structure of \Fig[e]{fig:1}. Below $5~\times~10^{-6}$~mbar and down to UHV [\Fig[d]{fig:2}], even larger \oneby{LP} patches are observed, while the remaining areas exhibit the stripes of \Fig[f]{fig:1}. Importantly, the process is reversible: STM confirms that the initial surface is regained when annealing back at high $p$\textsubscript{O$_2$}. A semi-ordered phase assigned to a transition state between the fishbone and the oblique structures appears around $10^{-4}$~mbar (not shown). 

Quantifying the \oneby{HP/LP} areal coverages at each step of the phase separation with STM [\Fig[a]{fig:3}] allows to determine the composition of the LP Mn-rich structures, as shown below. Note that because the following discussion focuses on relative cation compositions, which is the same on the two \oneby{HP} and \oneby{LP} phases, these will be henceforth referred to simply as \oneby{}, disregarding the change in atomic structure and oxygen content occurring below $2~\times~10^{-3}$ mbar O$_2$ at 700\degC{}. In \Fig[a]{fig:3}, full, black symbols represent the experiment of \Fig{fig:2}: As discussed, decreasing \muO{} produces increasingly larger \oneby{} coverages. The same trend is observed for starting Mn-richer surface compositions (gray), albeit with a smaller slope: Mn-richer surfaces form less of the Mn-poorer \oneby{} areas. Importantly, the same phases with the same quantitative coverages are observed under the same value of \muO{}, obtained from different combinations of temperature and $p$\textsubscript{O$_2$} [dashed oval in \Fig[a]{fig:3}]: This indicates that the observed surface structures are equilibrium phases, and that the phase separation is not kinetically limited. Recall that no phase separation occurs when starting from a \oneby{} surface (orange): The \oneby{} remains always monophase, hence the plot with \muO{} shows a constant area fraction of 100\%.

What drives the phase separation? One can rule out evaporation of cations (due to the reversibility), as well as cation diffusion to or from the bulk (at the employed conditions, cations can travel in bulk LSMO at most one atomic layer \cite{kubicek2014}): The phase separation thus occurs by mass transport across the surface. This is consistent with the fact that the average surface cation composition is conserved, as indicated by the reversibility of the process, and from the XPS data of \Fig[b]{fig:3}: These show that the intensity ratios of selected core level peaks have no trend with \muO{}. Given the sensitivity of XPS to the cation composition of monophase LSMO(110) structures \cite{franceschi2020atomically}, the absence of a trend indicates that the average surface composition is conserved. The proposed mechanism, sketched in \Fig[c]{fig:3}, is that Mn travels across the surface and exposes the very stable, Mn-poor \oneby{} phase, while enriching the remaining areas.  The LEIS data of \Fig[d]{fig:3}, showing that the monophase oblique phase is Mn-richer than the monophase fishbone, further support this scenario. Moreover, the stabilization mechanism of the Mn-rich phases at reducing conditions \emph{via} the increase in their Mn content is consistent with the lower Mn oxidation state displayed by Mn-richer HP structures \cite{franceschi2020atomically}.

\begin{figure}[t]
\includegraphics[width=\columnfigure]{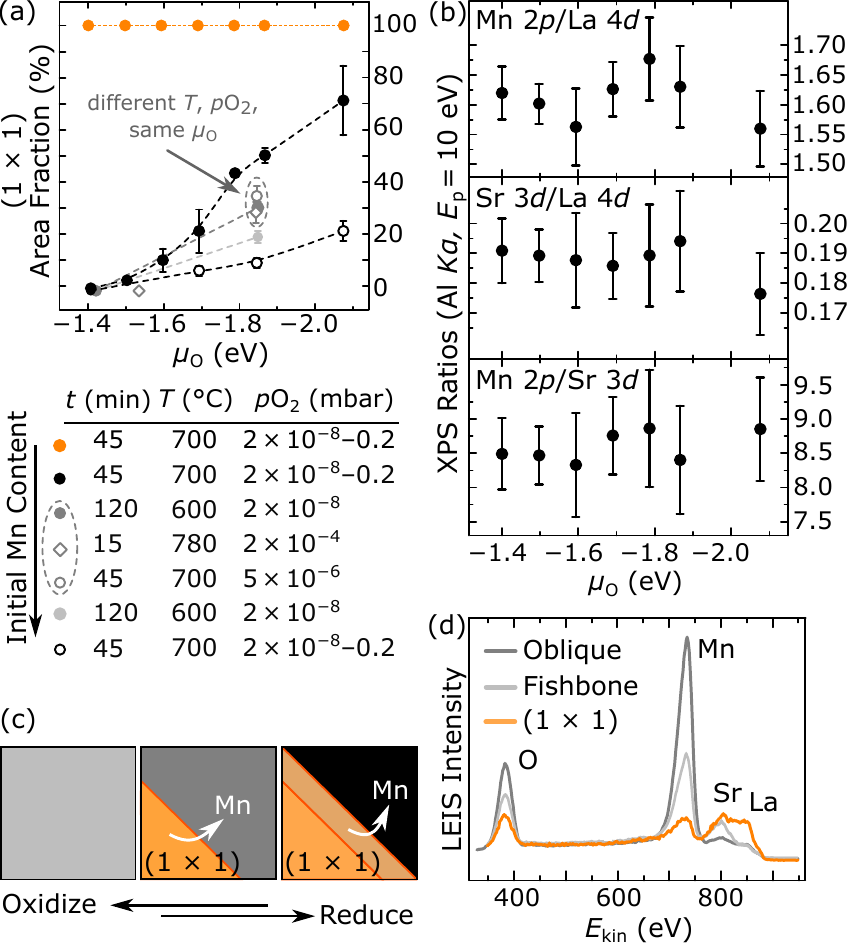}
\caption{\label{fig:3}Evaluation of surface compositions. (a) STM quantification of \oneby{} coverages during the phase separation of LSMO(110) surfaces, starting from different high-pressure phases. Each point averages over 10--12 images acquired on 300 $\times$ 300 nm$^2$ on different sample spots. Orange: starting \oneby{HP}. Full circles: starting fishbone (\Fig{fig:2}). Other symbols: starting surfaces between fishbone and \twoby{}. Dashed oval: different ($T$, $p$\textsubscript{O$_2$}) combinations, equivalent to the same \muO{}. The table indicates the annealing parameters for each data point. Lines are a guide for the eye. (b) XPS intensity ratios show no trend with \muO{} within the error bars (the data correspond to the experiment of \Fig{fig:2}, and are representative of all experiments). (c) Proposed mechanism: Mn travels across the surface to expose \oneby{} and form new Mn-richer structures, preserving the overall cation composition. (d) LEIS spectra of different monophase surfaces (see also Section S2).}
\end{figure}

Because of the current lack of knowledge on the surfaces of perovskite oxides, one cannot determine whether the \muO{}-dependent phase separation witnessed for LSMO(110) can be generalized to this material class. The only other perovskite oxide whose surfaces have been investigated in significant depth, SrTiO$_3$, shows a phase separation among stable surface structures with different A:B ratio \cite{riva2019epitaxial}, but not as a function of \muO{}. This is presumably because SrTiO$_3$ does not possess the same flexibility in the oxidation state of the B cation as LSMO \cite{kozakov2015valence,abbate1992controlled,saitoh1995electronic}, and because of its much smaller tendency to form oxygen vacancies. Materials with easily reducible cations are expected to behave similarly to LSMO(110): After all, structurally complex B-site-rich reconstructions that depend on the cation- and anion-composition seem to be a general trait of perovskite oxides \cite{andersen2018pauling,enterkin2010homologous,kolpak2008evolution}. Moreover, an AO termination was preferentially exposed at reducing conditions also in another manganite \cite{tselev2015surface}, and a similar effect was predicted for LSMO(001) \cite{hess2020polar}. 

\begin{figure}[t]
\includegraphics[width=\columnfigure]{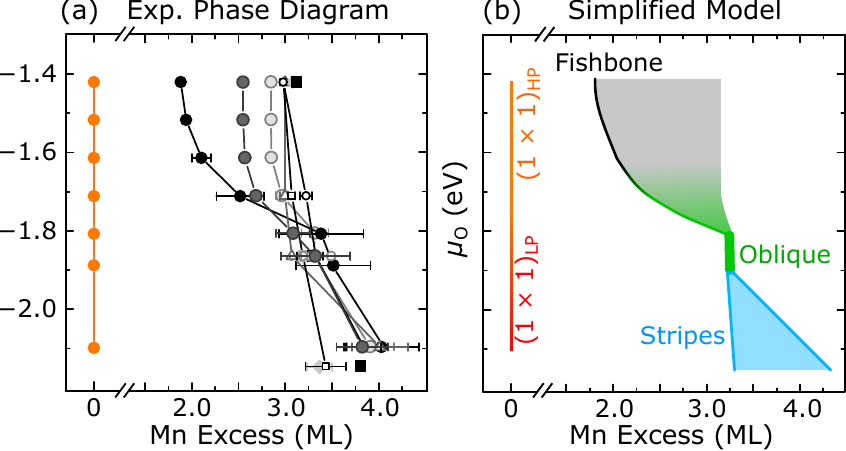}
\caption{\label{fig:4}Experimental 2D surface phase diagram of LSMO(110). (a) Each curve describes the evolution of a given high-pressure surface of LSMO(110) with decreasing \muO{}. (b) Corresponding sketch.}
\end{figure}

The remarkable diversity of the surface phases of perovskite oxides calls for methods capable of accessing and controlling their local surface properties as a function of different cation and anion compositions. Here, such a method is showcased for LSMO(110). As detailed below, it yields the quantitative cation compositions of Mn-rich, LP phases relative to \oneby{LP}, based on the measured coverages of the phase-separated structures [\Fig[a]{fig:3}], and on the known differences in Mn content of the HP phases \cite{franceschi2020atomically}. The outcome is shown in \Figs{fig:4}\hyperref[fig:4]{(a)} and \ref{fig:4}\hyperref[fig:4]{(b)}, displaying the experimental 2D surface phase diagram of LSMO(110) as a function of \muO{} and the cation composition, and the corresponding sketch, respectively. In \Fig[a]{fig:4}, each curve describes the evolution of a surface with given composition (horizontal axis) prepared at high $p$\textsubscript{O$_2$}, and then annealed at decreasing \muO{} (vertical axis). For instance, the left-most curve represents the evolution of the initially monophase \oneby{} surface: Its cation composition never changes, hence its evolution with \muO{} is a vertical line. The solid black circles represent the evolution of the initial fishbone surface, which phase-separates into \oneby{} areas and regions characterized by LP, Mn-richer structures (oblique or stripes, see \Fig{fig:2}). The other curves represent surfaces that have been prepared initially to exhibit a slightly Mn-richer structure [between the fishbone and the \twoby{} of \Fig[b, c]{fig:1}].

To derive the compositions of the LP phases formed during the phase separation, the number of cations on the whole surface was taken as constant at each annealing step, as supported by the reasonings above. This condition can be expressed as $S=\sum_{i}\theta_{i}s_{i} =\text{const}$, where $S$ is the density of Mn cations for a surface with multiple structures exposed, whose density of Mn cations and areal fractions are $s_{i}$ and $\theta_{i}$, respectively \cite{riva2019pushing}. Initially (at high pressure) the surface exposes only a Mn-rich, HP surface, and $S=s\textsubscript{HP}$. Referring this quantity to the density of Mn cations of the \oneby{}, $s$\textsubscript{\oneby{}}, one has $\Delta{s}=\Delta{s}\textsubscript{HP}=s\textsubscript{HP}-s\textsubscript{\oneby{}}$, which is known from \Fig{fig:1}: for the fishbone, $\Delta{s}\textsubscript{HP}=1.88(5)$~ML. By annealing at low pressure, the surface phase-separates into \oneby{} areas and new LP Mn-rich areas. Now, $\Delta{s}=\Delta{s}\textsubscript{LP}\,\theta\textsubscript{LP}$, where $\Delta{s}$\textsubscript{LP} is the (unknown) density of Mn cations on the new LP phase [referred to $s$\textsubscript{\oneby{}}], and $\theta$\textsubscript{LP} is the corresponding areal coverage (measured by STM). Since $\Delta{s}=\text{const}$ by assumption, one obtains that $\Delta{s}\textsubscript{LP}\,\theta\textsubscript{LP}=\Delta{s}\textsubscript{HP}$. The only unknown, $\Delta{s}\textsubscript{LP}$, can be derived and plotted in the phase diagram of \Fig[a]{fig:4}.

Note that most of the investigated structures belong to a common family in which a minor change of the cation composition induces a slight modification of the surface structure [consider the reconstructions between the fishbone and the \twoby{} \cite{franceschi2020atomically}, or the stripes discussed in Section S3]. As a result, they appear as `coexistence' regions in the phase diagram, not as lines [grey and blue in \Fig[b]{fig:4}, respectively].

In conclusion, this study demonstrates that the surfaces of perovskite oxides like LSMO(110) are very sensitive to the cation composition and \muO{}. Depending on the treatment conditions, the local structural properties and compositions may be affected drastically: Even if a monophase surface is obtained at some oxygen chemical potential, different annealing conditions might cause a phase separation and the emergence of new structures. This rich behavior agrees with Gibbs' phase rule, showcased here for the first time for the surface of a multielement oxide. The surfaces of other multielement oxides with easily reducible cations are expected to behave similarly, and to thus affect the performance of oxide-based-devices at different operating conditions. This work advances the needed knowledge and control of the surfaces of multielement oxides: It demonstrates how one can establish an experimental surface phase diagram over a wide range of cation and anion compositions, and how to quantify surface cation concentrations using phase separations.

\begin{acknowledgments}
This work was supported by the Vienna Science and Technology Fund (WWTF), the City of Vienna and Berndorf Privatstiftung through project MA 16-005. GF, UD, and MR were supported by FWF project F45-07 ``Functional Oxide Surfaces and Interfaces'' (FOXSI). MS was supported by FWF project F45-05 ``Functional Oxide Surfaces and Interfaces'' (FOXSI). GF acknowledges support by the Doctoral School TU-D of the TU Wien. 
\end{acknowledgments}

%\nolinenumbers
\providecommand{\enquote}[1]{#1}
% \bibliography{biblioLSMO2}
%merlin.mbs apsrev4-1.bst 2010-07-25 4.21a (PWD, AO, DPC) hacked
%Control: key (0)
%Control: author (8) initials jnrlst
%Control: editor formatted (1) identically to author
%Control: production of article title (0) allowed
%Control: page (1) range
%Control: year (1) truncated
%Control: production of eprint (0) enabled
%

\end{document}